\documentstyle[12pt]{article}
\def\nabstar#1{\nabla\kern-0.5pt\smash{\raise 4.5pt\hbox{$\ast$}}
               \kern-4.5pt_{#1}}

\def\drvstar#1{\partial\kern-0.5pt\smash{\raise 4.5pt\hbox{$\ast$}}
               \kern-5.0pt_{#1}}

\def\newline{\relax\ifhmode\null\hfil\break\else\nonhmodeerr@\newline\fi}
\def\frac#1#2{{#1\over#2}}
\def\text#1{{\hbox{\rm #1}}}
\def\flushpar{{\par \noindent}}

\newcommand{\beq}{\begin{equation}}
\newcommand{\eeq}{\end{equation}}
\newcommand{\bea}{\begin{eqnarray}}
\newcommand{\eea}{\end{eqnarray}}
\def\Id{ \mbox{1\hspace{-1.2mm}I} }
\def\BE{\begin{equation}}
\def\EE{\end{equation}}
\def\BA{\begin{eqnarray}}
\def\EA{\end{eqnarray}}
\def\BAN{\begin{eqnarray*}}
\def\EAN{\end{eqnarray*}}

\def\tr{\mbox{tr}}

\def\det{\mbox{det}}

\def\gm5{\gamma_5}

\def\Dcont{{\cal D}}

\def\anxL{{\cal A}_L(x)}

\input epsf.sty
\newdimen\psfigsize
\def\psfigure#1 #2 #3 #4 #5{
    \begin{figure}[tbh]
      \begin{center}
      \vbox{
        \null\vskip-0.2in\hskip#2
        \epsfxsize=#1
        \epsfbox{#4}
        \vskip -0.3in
        \caption {#5 \label{#3}}
        \vskip 0.0 true in plus 0.3 true in
      }
      \end{center}
   \end{figure}
}
\begin{document}
\thispagestyle{empty}
\begin{flushright}
NTUTH-00-101 \\
May 2000
\end{flushright}
\bigskip\bigskip\bigskip
\vskip 2.0truecm
\begin{center}
{\LARGE {Some remarks on the Ginsparg-Wilson fermion}}
\end{center}
\vskip 1.0truecm
\centerline{Ting-Wai Chiu}
\vskip5mm
\centerline{Department of Physics, National Taiwan University}
\centerline{Taipei, Taiwan 106, Republic of China.}
\centerline{\it E-mail : twchiu@phys.ntu.edu.tw}
\vskip 2cm
\bigskip \nopagebreak \begin{abstract}
\noindent

We note that Fujikawa's proposal of generalization of the
Ginsparg-Wilson relation is equivalent to setting
$ R = ( a \gamma_5 D )^{2k} $ in the original Ginsparg-Wilson relation
$ D \gamma_5 + \gamma_5 D = 2 a D R \gamma_5 D $.
An explicit realization of $ D $ follows from the Overlap construction.
The general properties of $ D $ are derived.
The chiral properties of these higher-order ( $ k > 0 $ )
realizations of Overlap Dirac operator are compared to those of the
Neuberger-Dirac operator ( $ k = 0 $ ),
in terms of the fermion propagator, the axial anomaly and
the fermion determinant in a background gauge field.
Our present results ( up to lattice size $ 16 \times 16 $ ) indicate that the 
chiral properties of the Neuberger-Dirac
operator are better than those of higher-order ones.


\end{abstract}
\vskip 1.5cm

\newpage\setcounter{page}1

\section{ Introduction }

During the last two years, it has become clear that the proper way
to formulate chiral fermions on the lattice is to impose the
exact chiral symmetry on the lattice, namely, the Ginsparg-Wilson
relation \cite{gwr}
\bea
\label{eq:gwr}
D \gm5 + \gm5 D = 2 a D R \gm5 D \ ,
\eea
where $ D $ is the lattice Dirac operator, $ a $ is the lattice
spacing, and $ R $ is a positive
definite Hermitian operator which commutes with $ \gm5 $.
Equation (\ref{eq:gwr}) should be regarded as a generalized chiral
symmetry which contains the usual chiral symmetry in the
continuum limit ( $ a \to 0 $ ). However, it should be noted
that this exact chiral symmetry ( $ R = 1 $ ) had been
existing in the Overlap formalism \cite{rn95,hn97:7}, even before the GW
relation was rediscovered. Therefore, unless one can explicitly construct
a GW Dirac operator $ D $ without using the Overlap, and such a $ D $
satisfies all physical requirements; otherwise it is unlikely that
the GW relation would turn out to be more fundamental than the Overlap.

Recently, Fujikawa \cite{fuji00:4} proposed a generalization of the
Ginsparg-Wilson relation as
\bea
\label{eq:gwr_of}
\gm5 ( \gm5 D ) + ( \gm5 D ) \gm5 =  2 a^{2k+1} ( \gm5 D )^{2k+2},
\hspace{4mm} k=0, 1, 2, \cdots
\eea
Multiplying both sides of (\ref{eq:gwr_of}) by $ \gm5 $, we obtain
\bea
\label{eq:gwr_f}
\gm5 D +  D \gm5 =  2 a D ( a \gm5 D )^{2k} \gm5 D ,
\hspace{4mm} k=0, 1, 2, \cdots
\eea
which is {\it equivalent} to the original GW relation (\ref{eq:gwr}) with
\bea
\label{eq:fuji_R}
R = ( a \gm5 D )^{2k}, \hspace{4mm} k=0, 1, 2, \cdots
\eea
It can be shown that $ R $ is Hermitian and commutes with $ \gm5 $
for any $ D $ which is $\gm5$-Hermitian and satisfies (\ref{eq:gwr_f})
[ the proof is below Eq. (\ref{eq:D52}) ].
The motivation of considering $ k > 0 $ in (\ref{eq:fuji_R}) is
to improve the chiral symmetry at {\it small} lattice spacings.
Further, Fujikawa has constructed a sequence ( $ k > 0 $ )
of these GW Dirac operators based on the
Neuberger-Dirac operator ( $ k = 0 $ ) \cite{hn97:7}.
However, the price one has to pay for the improved chiral symmetry
is a less localized $ D $, since in the limit $ k \to \infty $,
$ D $ must tend to a chirally symmetric and nonlocal $ D_c $, as a
consequence of the Nielson-Ninomiya no-go theorem \cite{no-go}.
Therefore, it is not clear whether one may have any advantages
in practice by considering $ k > 0 $ in (\ref{eq:fuji_R}).
Nevertheless, from a theoretical viewpoint, it is interesting
to see how one can construct a sequence ( $ k = 1, 2, \cdots $ )
of topologically proper $ D $ satisfying the GW relation (\ref{eq:gwr_f}),
in addition to the Neuberger-Dirac operator ( $ k=0 $ ).

In this paper, we examine several aspects of Fujikawa's proposal.
In section 2, we derive the analytical properties
of the GW Dirac operator satisfying (\ref{eq:gwr}) with
$ R = ( a \gm5 D )^{2k} $. Then, in section 3, we analyze the
construction of higher-order Overlap Dirac operators, and derive
their general properties.
In section 4, we compare the chiral properties of the higher-order
( $ k = 1, 2 $ ) Overlap Dirac operators to those of the Neuberger-Dirac
operator, by computing the fermion propagator, the axial anomaly and
the fermion determinant in two-dimensional background $ U(1) $ gauge fields.
Finally, we discuss and conclude in section 5.

\section{General analytical properties}

In this section, we begin with general considerations of
the GW relation, and then derive the analytical properties
of the GW Dirac operators satisfying (\ref{eq:gwr}) with
$ R = ( a \gm5 D )^{2k} $.

In general, one can assume that the lattice Dirac operator $ D $
satisfies the Ginsparg-Wilson relation in the form
\bea
\label{eq:ggw}
D \gm5 f(D) + g(D) \gm5 D = 0 \ ,
\eea
where $ f $ and $ g $ are any analytic functions. Then the
fermionic action $ {\cal A}_f = \bar\psi D \psi $ is invariant under the
global chiral transformation
\bea
\label{eq:gct}
\psi     \rightarrow \exp[ \theta \gm5 f(D) ] \ \psi       \\
\bar\psi \rightarrow \bar\psi \ \exp[ \theta g(D) \gm5 ]
\eea
where $ \theta $ is a global parameter.

If we set $ f(D) = \Id - a R D $ and $ g(D) = \Id - a D R $, then
(\ref{eq:ggw}) becomes
\bea
\label{eq:gwr_0}
D \gm5 + \gm5 D =  a D ( R \gm5 + \gm5 R ) D
\eea
where $ R $ is any operator.

Since the massless Dirac operator in continuum is chirally symmetric
( $ \Dcont \gm5 + \gm5 \Dcont = 0 $ ) and antihermitian
( $ \Dcont^{\dagger} = - \Dcont $ ), so it is
$\gm5$-Hermitian ( $ \Dcont^{\dagger} = \gm5 \Dcont \gm5 $ ).
Thus, we require that the lattice Dirac operator $ D $ also preserves
this symmetry at any lattice spacing, i.e.,
\bea
\label{eq:hermit}
D^{\dagger} = \gm5 D \gm5 \ .
\eea

Then multiplying (\ref{eq:gwr_0}) by $ \gm5 $ and using (\ref{eq:hermit}),
we obtain
\bea
\label{eq:DRD}
D^{\dagger} + D = a D^{\dagger} ( R + \gm5 R \gm5 ) D
                = a D ( R + \gm5 R \gm5 ) D^{\dagger} \ .
\eea
Evidently, only the part of $ R $ which commutes with $ \gm5 $
can enter (\ref{eq:DRD}). Recall that any operator $ R $ can be
decomposed into two parts as
\BAN
R  = \frac{1}{2} ( R + \gm5 R \gm5 ) + \frac{1}{2} ( R - \gm5 R \gm5 )
\EAN
where the first ( second ) term on r.h.s. commutes ( anticommutes )
with $ \gm5 $. Therefore, without loss, one can assume that
$ R $ commutes with $ \gm5 $. Thus, (\ref{eq:DRD}) becomes
\bea
\label{eq:DRD_1}
D^{\dagger} + D = 2 a D^{\dagger} R D
\eea
Taking the adjoint of (\ref{eq:DRD_1}), we immediately obtain
\bea
\label{eq:R_hermit}
R = R^{\dagger} \ .
\eea
Then (\ref{eq:gwr_0}) becomes the usual GW relation
\bea
\label{eq:gwr_o}
D \gm5 + \gm5 D =  2 a D R \gm5 D \ ,
\eea
where $ R $ is any Hermitian operator which commutes with $ \gm5 $.

Fujikawa's proposal \cite{fuji00:4} is equivalent to setting
\bea
\label{eq:fuji}
R = ( a \gm5 D )^{2k}, \hspace{4mm} k=0, 1, 2, \cdots
\eea
in the usual GW relation (\ref{eq:gwr_o}). Then (\ref{eq:gwr_o}) becomes
\bea
\label{eq:gwr_R}
D \gm5 + \gm5 D =  2 a D ( a \gm5 D )^{2k} \gm5 D \ .
\eea

It is obvious that $ R $ is Hermitian since $ D $ is $\gm5$-Hermitian.
Note that
\bea
\label{eq:D52}
\gm5 ( a \gm5 D )^2 = ( a \gm5 D )^2 \gm5
\eea
since
\BAN
& & \gm5 ( a \gm5 D )^2 \\
&=&  a^2  \gm5 ( \gm5 D + D \gm5 ) ( \gm5 D )
 - a^2 (\gm5 D ) \gm5 ( \gm5 D + D \gm5 ) + a^2 ( \gm5 D ) \gm5 ( D \gm5 ) \\
&=& 2 ( a \gm5 D )^{2k+2} - 2 ( a \gm5 D )^{2k+2} + ( a \gm5 D )^2 \gm5 \\
&=& ( a \gm5 D )^2 \gm5
\EAN
where (\ref{eq:gwr_R}) has been used in the second equality.
Then it follows that $ R = ( a \gm5 D )^{2k} $ commutes with $ \gm5 $.

Further, Eq. (\ref{eq:D52}) gives
\bea
\gm5 ( \gm5 D )^2 \gm5 = ( \gm5 D  )^2
\eea
which yields
\bea
\label{eq:DD}
D D^{\dagger} = D^{\dagger} D, \hspace{4mm} \mbox{ $ D $ is normal }
\eea
where (\ref{eq:hermit}) has been used.
Since $ D $ is normal, $ D $ and $ D^{\dagger} $ have common
eigenfunctions and their eigenvalues are either real or
come in complex conjugate pairs,
\bea
\label{eq:D_phi}
D \phi_s = \lambda_s \phi_s, \\
\label{eq:Dd_phi}
D^{\dagger} \phi_s = \lambda^{*} \phi_s,
\eea
and the eigenfunctions $ \{ \phi_s \} $ form a complete orthonormal set.
Then the general analytical properties of the eigenmodes as derived in
Ref. \cite{twc98:4} ( Eqs. (37), (38) and (41) in Ref. \cite{twc98:4} )
for $ R = 1/2 $ also hold for the $ R $ in (\ref{eq:fuji}).
For the sake of completeness, we outline the derivations as follows.
Writing $ R $ as
\bea
R = ( a \gm5 D )^{2k} = a^{2k} ( \gm5 D \gm5 D )^{k}
= a^{2k} ( D^{\dagger} D )^{k},
\eea
and applying Eq. (\ref{eq:DRD_1}) to $ \phi_s $, we obtain the
eigenvalue equation
\bea
\label{eq:eigen}
\lambda_s + \lambda_s^{*} = 2 a^{2k+1} (\lambda_s^{*})^{k+1} (\lambda_s)^{k+1}.
\eea
Multiplying both sides of (\ref{eq:eigen}) by
$ (\lambda_s)^k (\lambda_s^{*})^k $, we get
\bea
\label{eq:eigen_1}
(\lambda_s)^{k+1} (\lambda_s^{*})^{k} + (\lambda_s)^{k} (\lambda_s^{*})^{k+1}
= 2 a^{2k+1} (\lambda_s^{*})^{2k+1} (\lambda_s)^{2k+1}
\eea
which can be rewritten as
\bea
\label{eq:circle}
\left| \ (a \lambda_s)^{k+1} ( a \lambda_s^{*})^{k} - \frac{1}{2} \ \right|
= \frac{1}{2}
\eea
Thus the eigenvalues in the form
$ z = (a \lambda_s)^{k+1} ( a \lambda_s^{*})^{k} $
fall on the circle centered at $ 1/2 $ with radius $ 1/2 $.
The real eigenvalues ( if any ) of $ D $ are
at $ \lambda = 0 $ ( $ z = 0 $ ) and $ \lambda = a^{-1} $
( $ z = 1 $ ). Writing $ a \lambda_s = r \exp( i \theta ) $, we have
$ z = r^{2k+1} \exp( i \theta ) $.
Thus, for $ k = 0 $, the eigenvalues of $ D $
fall on the circle centered at $ 1/2 $ with radius $ 1/2 $,
while for $ k > 0 $, on the deformed circle stretched
symmetrically in $ \pm \hat{y} $ directions with fixed points at zero
and $ a^{-1} $, bounded inside the region $ 0 \le x \le a^{-1} $.

From (\ref{eq:hermit}) and (\ref{eq:Dd_phi}), we obtain
\bea
\label{eq:D_5_phi}
D \gm5 \phi_s = \lambda_s^{*} \gm5 \phi_s \ .
\eea
Multiplying both sides of (\ref{eq:D_5_phi}) by $ \phi_s^{\dagger} $
and using the adjoint of (\ref{eq:Dd_phi}), we get
\bea
\lambda_s \phi_s^{\dagger} \gm5 \phi_s =
\lambda_s^{*} \phi_s^{\dagger} \gm5 \phi_s
\eea
This implies that the chirality of any complex eigenmode is zero,
\bea
\label{eq:complex}
\chi_s \equiv \phi_s^{\dagger} \gm5 \phi_s = 0 \hspace{4mm}
\mbox{ if } \lambda_s \ne \lambda_s^{*}.
\eea
If $ \lambda_s $ is real ( zero or $ a^{-1} $ ),
then Eqs. (\ref{eq:D_5_phi}) and (\ref{eq:D_phi}) imply that
$ \phi_s $ has definite chirality $ +1 $ or $ -1 $ :
\bea
\label{eq:real}
\gm5 \phi_s = \pm \phi_s, \hspace{4mm}
\mbox{ if } \lambda_s = \lambda_s^{*}.
\eea
A useful property of chirality is that the total chirality
of all eigenmodes must vanish,
\bea
\sum_s \chi_s &=& \sum_s \phi_s^{\dagger} \gamma_5 \phi_s  \nonumber \\
&=& \sum_s \sum_x \sum_{\alpha} \sum_{\beta}
[\phi_s^{\alpha}(x)]^{*} \gamma_5^{\alpha \beta} \phi_s^{\beta}(x) \nonumber \\
&=& \sum_{\alpha} \sum_{\beta} \gamma_5^{\alpha \beta}
    \delta_{\alpha \beta} = 0
\label{eq:chisum}
\eea
where the completeness relation 
\beq
\sum_x \sum_s [\phi_s^{\alpha}(x)]^{*} \phi_s^{\beta}(x)
= \delta^{\alpha \beta}
\eeq
has been used. Since the chirality of any complex eigenmode is zero,
then Eq. (\ref{eq:chisum}) gives the chirality sum rule \cite{twc98:4}
for real eigenmodes,
\bea
 N_{+} + n_{+} = N_{-} + n_{-}
\label{eq:chi_sum_rule}
\eea
where $ n_{+} ( n_{-} ) $ denotes the number of zero modes of
positive ( negative ) chirality, and $ N_{+} ( N_{-} ) $ the number
of nonzero real ( $ a^{-1} $ ) eigenmodes of positive ( negative )
chirality. Then we immediately see that any zero mode must be
accompanied by a real ( $ a^{-1} $ ) eigenmode with opposite chirality,
and the index of $ D $ is
\bea
\label{eq:index_D}
\mbox{index}(D) \equiv n_{-} - n_{+} = - ( N_{-} - N_{+} ) \ .
\eea
It should be emphasized that the chiral properties (\ref{eq:real}),
(\ref{eq:complex}) and the chirality sum rule (\ref{eq:chi_sum_rule})
hold for any normal $ D $ satisfying the $\gm5$-Hermiticity,
as shown in Ref. \cite{twc98:4}. However, in nontrivial gauge
backgrounds, whether $ D $ possesses any zero modes or not
relies on the topological characteristics \cite{twc99:11} of $ D $,
which cannot be guaranteed by the conditions such as the locality,
free of species doublings, correct continuum behavior,
$\gm5$-Hermiticity and the GW relation.

\section{Higher-order realization of Overlap Dirac operator}

Now the task is to construct a topologically proper
$ D $ which is local, free of species doubling, $\gm5$-Hermitian,
having correct continuum behavior, and satisfies the GW relation
(\ref{eq:gwr_f}). So far, the only viable way to construct a
topologically proper $ D $ is the Overlap,
\bea
\label{eq:overlap}
D = \frac{1}{2a} ( \Id + \gamma_5 \epsilon ), \hspace{4mm} \epsilon^2 = \Id
\eea
which satisfies the GW relation (\ref{eq:gwr}) with $ R = 1 $.
There are many different ways to implement the Hermitian
$ \epsilon $ in (\ref{eq:overlap}).
However, it is required to be able to capture the topology of the gauge
background. That means, one-half of the difference of the numbers
( $ h_\pm $ ) of positive ( $ +1 $ ) and negative ( $ -1 $ )
eigenvalues of $ \epsilon $ is equal to the background
topological charge $ Q $,
\bea
\label{eq:index_Q}
  \sum_x \tr ( a \gm5 D(x,x) ) = \frac{1}{2} \sum_x \tr( \epsilon )
= \frac{1}{2} ( h_{+} - h_{-} ) = Q \ ,
\eea
where $ \tr $ denotes the trace over the Dirac and color space.
Otherwise, the axial anomaly of $ D $ cannot agree with the
topological charge density in a nontrivial gauge background.
Henceforth, we shall regard any lattice Dirac operator which
is constructed through the general form of Overlap (\ref{eq:overlap})
as a realization of the Overlap Dirac operator.

An explicit realization of $ \epsilon $ in
(\ref{eq:overlap}) is the Neuberger-Dirac operator \cite{hn97:7}
with
\bea
\label{eq:eh}
\epsilon = \frac{ H_w } { \sqrt{ H_w^2 } }
\eea
where
\bea
\label{eq:Hw}
H_w &=& \gm5 ( D_W - m_0 a^{-1} ), \hspace{4mm} 0 < m_0 < 2 r_w \ , \\
\label{eq:D_W}
D_W &=& \gamma_\mu t_\mu + W, \hspace{4mm}
    \mbox{ $ D_W $ : massless Wilson-Dirac operator }, \\
\label{eq:tmu}
t_\mu (x,y) &=& \frac{1}{2a} \ [   U_{\mu}(x) \delta_{x+\hat\mu,y}
                       - U_{\mu}^{\dagger}(y) \delta_{x-\hat\mu,y} ] \ , \\
\label{eq:wilson}
W(x,y) &=&  \frac{r_w}{2a} \sum_\mu \left[ 2 \delta_{x,y}
                     - U_{\mu}(x) \delta_{x+\hat\mu,y}
                     - U_{\mu}^{\dagger}(y) \delta_{x-\hat\mu,y} \right] \ .
\eea
( The Wilson parameter $ r_w $ is usually set to one. )
\BAN
\gamma_\mu = \left( \begin{array}{cc}
                            0                &  \sigma_\mu    \\
                    \sigma_\mu^{\dagger}     &       0
                    \end{array}  \right) \ ,
\EAN
and
\BAN
\sigma_\mu \sigma_\nu^{\dagger} + \sigma_{\nu} \sigma_\mu^{\dagger} =
2 \delta_{\mu \nu} \ .
\EAN
Note that the parameter $ m_0 $ plays a crucial role in
detecting the topology of the gauge background.

Now the problem is how to generalize this construction
to $ D $ satisfying the GW relation with
$ R = ( a \gm5 D )^{2k} $ for $ k > 0 $.
We can multiply both sides of the GW relation (\ref{eq:gwr})
by $ R $ and redefine $ D' = R D $, then we have
\bea
\label{eq:gwr_1}
D' \gm5 + \gm5 D' = 2 D' \gm5  D'
\eea
where $ D' = R D $ is $\gm5$-hermitian since
\BAN
D'^{\dagger} = D^{\dagger} R = \gm5 D \gm5 R = \gm5 D R \gm5 = R \gm5 D \gm5
= \gm5 R D \gm5 = \gm5 D' \gm5 \ . 
\EAN  
Now (\ref{eq:gwr_1}) is in the same form of the GW relation with $ R = 1 $.
Thus, one can construct $ D' $ in the same way as the Overlap
\bea
\label{eq:D'}
D' = R D = \frac{1}{2a} ( \Id + \gm5 \epsilon ), \hspace{4mm}
\epsilon^2 = \Id \ ,
\eea
provided that a proper realization of $ \epsilon $ can be obtained.
An explicit construction based on the Neuberger-Dirac operator
( $ k = 0 $ ) has been generalized to higher-orders
( $ k > 0 $ ) by Fujikawa \cite{fuji00:4}.

In the following, we formulate Fujikawa's construction in a more
transparent way. Using $ R \gm5 = \gm5 R $, we can write
\bea
D' = R D = ( a \gm5 D )^{2k} D = ( a \gm5 D )^{2k} \gm5 \gm5 D
   = a^{-1} \gm5 ( a \gm5 D )^{2k+1} \ ,
\eea
which yields
\bea
\label{eq:DD'}
D = a^{-1} \gm5 ( a \gm5 D')^{1/(2k+1)} \ ,
\eea
where the $(2k+1)$-th real root of the
Hermitian operator $ a \gm5 D ' $ is assumed.

Then (\ref{eq:DD'}) suggests that if the $ \epsilon $ in (\ref{eq:D'})
is expressed in terms of a Hermitian operator $ H $,
\bea
\label{eq:H}
\epsilon = \frac{H}{\sqrt{H^2}} \ ,
\eea
then $ H $ is required to be proportional to
$ ( \gamma_\mu \Dcont_\mu )^{2k+1} $ plus higher-order terms
in the continuum limit such that $ D $ behaves as
$ \gamma_\mu \Dcont_\mu $ after
taking the $ ( 2k+1 ) $-th root in (\ref{eq:DD'}).
Thus, $ H $ must contain the term $ ( \gamma_\mu t_\mu )^{2k+1} $,
where $ \gamma_\mu t_\mu $ is the naive lattice fermion operator
defined in (\ref{eq:tmu}). Then additional terms must be required
in order to remove the species doublers in the term
$ ( \gamma_\mu t_\mu )^{2k+1} $. So, we add the Wilson term
to the $(2k+1)$-th power, i.e., $ W^{2k+1} $.
Finally, a negative mass term
$ -(m_0 a^{-1} )^{2k+1} $ is inserted such that $ \epsilon $
is able to detect the topological charge $ Q $ of the
gauge background, i.e.,
\bea
\label{eq:HQ}
\frac{1}{2} \sum_x \tr \left( \frac{H}{\sqrt{H^2}} \right) = Q \ .
\eea
Putting all these terms together, we have
\bea
\label{eq:Hk}
H = \gm5 [ (\gamma_\mu t_\mu)^{2k+1} + W^{2k+1} - ( m_0 a^{-1} )^{2k+1} ] \ ,
\eea
which, at $ k=0 $, reduces to the $ H_w $ in Eq. (\ref{eq:Hw}).
Then (\ref{eq:DD'}) can be rewritten as
\bea
\label{eq:Dk}
D = a^{-1} \left( \frac{1}{2} \right)^{1/(2k+1)}
    \gm5 \left( \gm5 + \frac{H}{\sqrt{H^2}} \right )^{1/(2k+1)},
\eea
where $ H $ is defined in (\ref{eq:Hk}).
This is the higher-order realization of Overlap Dirac operator,
as constructed by Fujikawa \cite{fuji00:4}.

Next we derive some general properties of the Overlap Dirac operator
$ D $ defined in (\ref{eq:Dk}).

The fermion propagator $ S_F(x,y) $ is defined by
\BAN
S_F(x,y) = \frac{1}{Z} \int \prod_z d \bar{\psi}(z) d \psi(z)
           \ \text{e}^{- \bar\psi D \psi } \ \psi(x) \bar{\psi}(y)
\EAN
where
\BAN
Z = \int \prod_z d \bar{\psi}(z) d\psi(z) \text{e}^{- \bar\psi D \psi}
\EAN
In a background gauge field of zero topological charge ( $ Q = 0 $ ),
the fermion propagator is
\begin{eqnarray}
S_F(x,y) = D^{-1}(x,y)  \ .
\label{eq:SF}
\end{eqnarray}

In the naive continuum limit with $ r_w a^{-1} $
and $ m_0 a^{-1} $ kept finite, the ( free ) fermion propagator
in momentum space can be obtained after some straightforward algebras,
\beq
\tilde{S_F} (p) = a \gm5 \left[ \gm5 \left( \Id +
    \frac{1}{ (\gamma_\mu t_\mu)^{2k+1} } T(p) \right) \right]^{1/(2k+1)}
\label{eq:SF0}
\eeq
where
\bea
\label{eq:tmup}
t_\mu &=& i a^{-1} \sin( p_\mu a ) \ , \\
\label{eq:t2}
t^2   &=& a^{-2} \sum_\mu \sin^2( p_\mu a ) \ , \\
\label{eq:Wp}
W(p)  &=& r_w a^{-1} \sum_\mu \ [ 1 - \cos( p_\mu a ) ] \ ,     \\
\label{eq:up}
u(p)  &=&  [W(p)]^{2k+1} - (m_0 a^{-1})^{2k+1} \ , \\
\label{eq:Np}
N(p)  &=& \sqrt{(t^2)^{2k+1}+ u^2(p) } \ , \\
\label{eq:Tp}
T(p)  &=&  N(p) - u(p) \ .
\eea

Around $ p \simeq 0 $, $ t_\mu \simeq i p_\mu $,
$ T(p) \simeq  2 ( m_0 a^{-1} )^{2k+1} $,
and the fermion propagator becomes
\bea
\tilde{S_F} (p) & \simeq &
  a \gm5 \left[ \gm5 \left( \Id + \frac{1}{ ( i \gamma_\mu p_\mu )^{2k+1} }
           2 ( m_0 a^{-1} )^{2k+1} \right) \right]^{1/(2k+1)} \\
& \simeq & 2^{1/(2k+1)} m_0 \ \frac{1}{ i \gamma_\mu p_\mu }
    + a + a ( 1 - \delta_{k,0} ) \Sigma_k ( a p )
\label{eq:SF_zero}
\eea
where except for $ k = 0 $, a momentum-dependent term denoted by
$ a \Sigma_k ( a p ) $ is present in the scalar part,
which may lead to additive mass renormalization.
Evidently, we have to fix the value of $ m_0 $ to
\bea
\label{eq:m0}
 m_0 = \left( \frac{1}{2} \right)^{1/(2k+1)}
\eea
such that in the limit ( $ a \to 0 $ ) the fermion propagator
(\ref{eq:SF_zero}) agrees with the continuum propagator.

For $ m_0 \in (0, 2 r_w ) $, on a $d$-dimensional lattice ( $d$ = even ),
at any one the $ 2^d - 1 $ corners of the Brillouin zone [  i.e.,
$ a p = ( \pi, 0, \cdots, 0 ) $, $ ( 0, \pi, \cdots, 0 ) $, $ \cdots $,
$ ( \pi, \pi, \cdots, \pi ) $ ], we have $ N(p) = u(p) > 0 $,
thus $ T(p) = u(p) - u(p) = 0 $, and all doubled modes are
decoupled from the fermion propagator (\ref{eq:SF0}).

In general, we consider an arbitrary value of $ m_0 $.
At the origin ( $ p = 0 $ ) and the $ 2^d - 1 $ corners of the
Brillouin zone, we have $ \sin( p_\mu a ) = 0 $,
so $ T(p) $ becomes
\bea
\label{eq:Tp0}
T(p) = | u(p) | - u(p) \ ,
\eea
where the possible values of $ u(p) $ are :
\BAN
u(p) &=& -( m_0 a^{-1} )^{2k+1},  \\
     & & ( 2 r_w a^{-1} )^{2k+1} -( m_0 a^{-1} )^{2k+1} ,    \\
     & & ( 4 r_w a^{-1} )^{2k+1} -( m_0 a^{-1} )^{2k+1} ,    \\
     & & \cdots                                              \\
     & & \cdots                                              \\
     & & ( 2 d r_w a^{-1} )^{2k+1} -( m_0 a^{-1} )^{2k+1} \ .
\EAN
Here the first value of $ u(p) $ corresponds to all components of
$ p $ equal to zero, the second value to one of components equal
to $ \pi/a $, and so on, and the last value to all components
equal to $ \pi/a $.
Note that
\BAN
& & ( 2 n  r_w a^{-1} )^{2k+1} -( m_0 a^{-1} )^{2k+1}  \\
&=& ( 2 n r_w a^{-1} - m_0 a^{-1} )
    [ ( 2 n  r_w a^{-1} )^{2k}  + \cdots + ( m_0 a^{-1} )^{2k} ],
  \hspace{2mm} n = 0, 1, \cdots, d.
\EAN
Therefore the sign of $ u(p) $ is independent of the order $ k $.
From Eq. (\ref{eq:Tp0}), we see that if $ u(p) \ge 0 $, then $ T(p)=0 $,
and this doubled mode is decoupled from the fermion propagator
(\ref{eq:SF0}) for any order $ k $. Since the chiral charge of a doubled
mode is equal to $ (-1)^n $, where $ n $ is the number of
momentum components equal to $ \pi/a $, then the total
chiral charge $ Q_5 $ of all massless ( primary and doubled )
fermion modes contributing to the fermion propagator (\ref{eq:SF0})
can be determined. Then $ \mbox{index} ( D ) = Q_5 Q $
for a gauge background with topological charge $ Q $.
Thus, in the naive continuum limit, the index of $ D $
( as a function of $ m_0 $ ) is independent of the order $ k $,
same as the index of the Neuberger-Dirac operator ( $ k = 0 $ ),
which has been determined in Ref. \cite{twc98:10a}. Explicitly,
\beq
\label{eq:index_sym}
\mbox{index}[D(m_0)] = \left\{  \begin{array}{ll}
\frac{ (-1)^{n+1} (d-1)! }{ (d-n)! \ (n-1)! } \ Q \ ,
     & \mbox{ \ $ 2(n-1) r_w  < m_0 < 2 n r_w $} \\
$ $  & \mbox{ \ \  \ for $ n=1,\cdots,d $ ; }            \\
 0,  & \mbox{ otherwise.   }  \\
       \end{array}
       \right.
\eeq
In particular, for $ d = 4 $,
\beq
\label{eq:index_4d}
\mbox{index}[D(m_0)] = \left\{  \begin{array}{ll}
\ \ \ \ Q, &      0 < m_0 < 2 r_w \ ,   \\
      -3Q, &  2 r_w < m_0 < 4 r_w \ ,   \\
 \ \   3Q, &  4 r_w < m_0 < 6 r_w \ ,   \\
 \     -Q, &  6 r_w < m_0 < 8 r_w \ ,   \\
\ \ \ \ 0, & \mbox{ otherwise.   }      \\
                  \end{array}
              \right.
\eeq

For the Neuberger-Dirac operator ( $ k = 0 $ ),
there exists an exact discrete symmetry of the index on {\it any}
finite lattice with even number of sites in each dimension \cite{twc98:10a}
\bea
\label{eq:index_ref_sym}
\mbox{index}[ D(m_0) ] = - \mbox{index} [ D( 2 d r_w - m_0 ) ], \hspace{4mm}
\mbox{ for } k = 0,
\eea
which holds for any background gauge configuration. However,
for higher-order Overlap Dirac operators ( $ k > 0 $ ),
this discrete symmetry is {\it not} exact on a
finite lattice; only in the naive continuum limit,
this discrete symmetry can be realized as in (\ref{eq:index_sym}).
Nevertheless, at $ m_0 = d r_w $, it can be
shown that the index is exactly zero for all $ k $,
\bea
\label{eq:index_zero}
\mbox{index}[ D( m_0 = d r_w ) ] = 0, \hspace{4mm} \mbox{ for all } k \ge 0 \ ,
\eea
on any finite lattice with even number of sites in each dimension,
and for any background gauge configuration.

\section{ Tests }

In this section, we compare the chiral properties of the higher-order
Overlap Dirac operators ( $ k = 1, 2 $ ) to those of the Neuberger-Dirac
operator ( $ m_0 = 1 $ and $ R = 1/2 $ ), by computing the fermion
propagator, the axial anomaly and the fermion determinant
in two-dimensional background $ U(1) $ gauge fields.
Our notations for the two-dimensional background
gauge field are the same as those in Ref. \cite{twc98:4}
( Eqs. (7)-(11) in Ref. \cite{twc98:4} ).

Note that the Neuberger-Dirac operator is conventionally written as
\bea
\label{eq:Dh}
D = a^{-1} ( \Id + V ) = a^{-1} ( \Id + \gm5 \frac{H_w}{\sqrt{H_w^2}} ),
\hspace{4mm}  m_0 = 1;
\eea
which satisfies the GW relation with $ R = 1/2 $.
However, the zeroth order ( $ k = 0 $ ) Overlap Dirac operator is
\bea
\label{eq:k0}
D = \frac{1}{2a} ( \Id + \gm5 \frac{H_w}{\sqrt{H_w^2}} ),
\hspace{4mm}  m_0 = 1/2;
\eea
which satisfies the GW relation with $ R = 1 $.
In the following, we shall use the Neuberger-Dirac operator (\ref{eq:Dh})
in place of the zeroth order Overlap (\ref{eq:k0}). All numerical results
for the $ k=0 $ case are obtained using the Neuberger-Dirac operator
(\ref{eq:Dh}) rather than (\ref{eq:k0}).


First of all, we checked that the eigenvalues of a higher-order
( $ k = 1, 2 $ ) Overlap Dirac operator fall on the deformed circle
which is stretched symmetrically in $ \pm \hat{y} $ directions.
In a nontrivial gauge background,
the real eigenmodes ( zero and $ a^{-1} $ ) have definite chirality
and satisfy the chirality sum rule (\ref{eq:chi_sum_rule}),
and each complex eigenmode has zero chirality (\ref{eq:complex}).
The Atiyah-Singer index theorem ( $ n_{-} - n_{+} = Q $ )
is satisfied in all cases ( $ k = 0, 1, 2 $ ) for
gauge configurations fulfiling the topological bound \cite{twc99:11}
\bea
\label{eq:top_bd}
a^2 | \bar\rho(x) | < \epsilon_1 \simeq 0.28 \hspace{4mm} \forall x
\eea
where $ a^2 \bar\rho(x) $ is the topological charge inside the unit
square of area $ a^2 $ centered at $ x $,
\bea
\label{eq:rho_bar}
\bar\rho(x) = \frac{1}{a^2} \int_{x_1 -a/2}^{x_1+a/2} dy_1
              \int_{x_2 -a/2}^{x_2+a/2} dy_2 \ \frac{1}{2\pi} F_{12} (y) \ .
\eea
The value of $ m_0 $ in the higher-order ( $ k > 0 $ ) Overlap Dirac operator
is fixed according to Eq. (\ref{eq:m0}), $ m_0 = 2^{-1/(2k+1)} $,
while $ m_0 = 1 $ for the Neuberger-Dirac operator.

\subsection{Fermion propagator}

In the following, we first compute the free fermion propagator
on a two dimensional lattice, for
the Neuberger-Dirac operator and higher-order ( $ k = 1, 2 $ )
Overlap Dirac operators respectively. We compare them to
the exact solution of the massless fermion propagator on the torus.
Then we turn on a background gauge field to examine the behaviors
of the scalar part $ S_0 $ and the pseudoscalar part $ S_5 $
in the higher order ( $ k = 1, 2 $ ) fermion propagators.

In general, the free fermion propagator can be written as
\bea
\label{eq:SFx}
S_F (x) = S_0 (x) + \gamma_\mu S_\mu (x) \ ,
\eea
where $ S_0 (x) = 0 $ for the massless fermion in continuum;
and $ S_0 (x) = ( a/2  ) \delta_{x,0} $ for the Neuberger-Dirac operator.

First, we examine the $ S_\mu (x) $ components in (\ref{eq:SFx}).
In Table 1, we list the component $ S_1(x) $
along the diagonal ( $ x_1 = x_2 $ ) of a $ 16 \times 16 $ lattice
with antiperiodic boundary conditions. One of the end points
of the propagator is fixed at the origin, while the other
end point is located at a site along the diagonal.
Note that, by symmetry, $ S_2(x) = S_1 (x) $ along the diagonal.
From the data in Table 1, we immediately see that
in all cases ( Neuberger, $ k = 1, 2 $ ),
$ S_1(x) $ agrees very well with the exact solution on the torus.
In general, all $ S_{\mu} $ components of the free fermion propagators
are in good agreement with the exact solution for any $ x = ( x_1, x_2 ) $.

Next, we examine the scalar part $ S_0 (x) $ in the free fermion
propagator of the higher-order ( $ k = 1, 2 $ )
Overlap Dirac operators (\ref{eq:Dk}).
In Table 2, we list $ S_0 (x) $ along the diagonal of the
$ 16 \times 16 $ lattice with antiperiodic boundary conditions.
From the data in Table 2, we see that $ S_0 (x) $ is local for
both $ k=1 $ and $ k=2 $.
However, we note that $ S_0 (x) $ in the second order ( $ k = 2 $ ) case
is less localized than that of the first order ( $ k = 1 $ ), as expected.
It seems that $ | S_0(x) | $ in both ( $ k = 1, 2 $ ) orders can be
fitted by an exponentially decay function for $ 0 < | x | < 8 \sqrt{2} $.

{\footnotesize
\begin{table}
\begin{center}
\begin{tabular}{|c|c|c|c|c|c|}
\hline
$ x_1 $ & $ x_2 $ &  Exact &  Neuberger  &  $ k = 1 $  &  $ k = 2 $ \\
\hline
\hline
0.0  &  0.0  &  0.0000  &  0.0000  &  0.0000  &  0.0000 \\
\hline
1.0  &  1.0  &  0.0796  &  0.0689  &  0.0894  &  0.0885  \\
\hline
2.0  &  2.0  &  0.0396  &  0.0374  &  0.0423  &  0.0449  \\
\hline
3.0  &  3.0  &  0.0259  &  0.0249  &  0.0226  &  0.0212   \\
\hline
4.0  &  4.0  &  0.0184  &  0.0179  &  0.0177  &  0.0179    \\
\hline
5.0  &  5.0  &  0.0131  &  0.0127  &  0.0128  &  0.0126    \\
\hline
6.0  &  6.0  &  $ 8.59 \times 10^{-3} $  &  $ 8.32 \times 10^{-3} $
&  $ 8.32 \times 10^{-3} $ &  $  8.38 \times 10^{-3} $   \\
\hline
7.0  &  7.0  &  $ 4.28 \times 10^{-3} $  &  $ 4.13 \times 10^{-3} $
&  $ 4.10 \times 10^{-3} $ &  $ 4.17 \times 10^{-3} $   \\
\hline
8.0  &  8.0  &  0.0000  &  0.0000  &  0.0000  &  0.0000     \\
\hline
9.0  &  9.0  & $ -4.28 \times 10^{-3} $  & $ -4.13 \times 10^{-3} $
& $ -4.10 \times 10^{-3} $ & $ -4.17 \times 10^{-3} $   \\
\hline
10.0 & 10.0  & $ -8.59 \times 10^{-3} $ & $ -8.32 \times 10^{-3} $
& $ -8.32 \times 10^{-3} $ & $ -8.38 \times 10^{-3} $   \\
\hline
11.0 & 11.0  & -0.0131  & -0.0127  & -0.0128  & -0.0126    \\
\hline
12.0 & 12.0  & -0.0184  & -0.0179  & -0.0177  & -0.0179    \\
\hline
13.0 & 13.0  & -0.0259  & -0.0249  & -0.0226  & -0.0212    \\
\hline
14.0 & 14.0  & -0.0396  & -0.0374  & -0.0423  & -0.0449     \\
\hline
15.0 & 15.0  & -0.0796  & -0.0689  & -0.0894  & -0.0885    \\
\hline
\end{tabular}
\end{center}
\caption{The free fermion propagators
on a $ 16 \times 16 $ lattice with antiperiodic boundary conditions.
One of the end points of the propagator is fixed at the origin, while the
other end point is at one of the sites along the diagonal ( $ x_1 = x_2 $ ).
}
\label{table:1}
\end{table}
}

{\footnotesize
\begin{table}
\begin{center}
\begin{tabular}{|c|c|c|c|}
\hline
$ x_1 $ & $ x_2 $ &  $ S_0(x) $, $ k = 1 $  & $ S_0(x) $,  $ k = 2 $  \\
\hline
\hline
0.0  &   0.0   &   0.9170  &  0.8950                                \\
\hline
1.0  &   1.0   &  -0.0488  & -0.0630                                \\
\hline
2.0  &   2.0   &  $ -9.45 \times 10^{-3} $ &   -0.0085              \\
\hline
3.0  &   3.0   &  $ -4.41 \times 10^{-4} $ &  $  3.30 \times 10^{-3} $   \\
\hline
4.0  &   4.0   &  $ 2.33 \times 10^{-7} $  &  $  1.56 \times 10^{-3} $   \\
\hline
5.0  &   5.0   &  $ 1.63 \times 10^{-4} $  &  $  5.96 \times 10^{-4} $   \\
\hline
6.0  &   6.0   &  $ 1.12 \times 10^{-4} $  &  $ -2.52 \times 10^{-4} $   \\
\hline
7.0  &   7.0   &  $ 3.47 \times 10^{-5} $  &  $ -1.72 \times 10^{-4} $   \\
\hline
8.0  &   8.0   &   0.0000  &  0.0000                                \\
\hline
9.0  &   9.0   &  $ 3.47 \times 10^{-5} $  &  $ -1.72 \times 10^{-4} $   \\
\hline
10.0 &  10.0   &  $ 1.12 \times 10^{-4} $  &  $ -2.52 \times 10^{-4} $   \\
\hline
11.0 &  11.0   &  $ 1.63 \times 10^{-4} $  &  $  5.96 \times 10^{-4} $   \\
\hline
12.0 &  12.0   &  $ 2.33 \times 10^{-7} $  &  $  1.56 \times 10^{-3} $   \\
\hline
13.0 &  13.0   &  $-4.41 \times 10^{-4} $  &  $  3.30 \times 10^{-3} $   \\
\hline
14.0 &  14.0   &  $-9.45 \times 10^{-3} $  &     -0.0085                 \\
\hline
15.0 &  15.0   &  -0.0488                  &     -0.0630                 \\
\hline
\end{tabular}
\end{center}
\caption{The scalar part $ S_0 (x) $ in the free fermion propagator
of the higher-order Overlap Dirac operators, along the diagonal of a
$ 16 \times 16 $ lattice with antiperiodic boundary conditions. }
\label{table:2}
\end{table}
}

In a background gauge field, the fermion propagator can be written as
\bea
\label{eq:SFXA}
S_F (x,y) = \left( \begin{array}{cc}
                      S_0(x,y) + S_5(x,y)    &  S_R(x,y)    \\
                      S_L(x,y)               &  S_0(x,y) - S_5(x,y)
                    \end{array}  \right) \ ,
\eea
where $ S_0 (x,y) = S_5(x,y) = 0 $ for the massless fermion in continuum;
and $ S_0 (x,y) = ( a/2  ) \delta_{x,y} $ and $ S_5 (x,y) = 0 $ for the
Neuberger-Dirac operator. However, for higher-order ( $ k > 0 $ )
Overlap Dirac operators, both $ S_0(x,y) $ and $ S_5 (x,y) $ are not
proportional to $ \delta_{x,y} $.
If $ S_0 (x,y) $ ( $ S_5 (x,y) $ ) turns out to be nonlocal, then
it would cause additive mass renormalization
and the poles in the fermion propagator will be shifted accordingly.

Now we turn on a background $ U(1) $ gauge field with parameters
( $ h_1 = 0.1 $, $ h_2 = 0.2 $, $ A_1^{(0)} = 0.3 $, $ A_2^{(0)} = 0.4 $
and $ n_1 = n_2 = 1 $, as defined in Eqs. (7) and (8) in
Ref. \cite{twc98:4} ). Then we examine the behaviors of $ S_0(x,y) $
and $ S_5(x,y) $ for the higher-order ( $ k = 1, 2 $ )
Overlap Dirac operators.
We find that both $ S_0(x,y) $ and $ S_5(x,y) $
are local in the higher-order ( $ k = 1, 2 $ ) fermion propagators.
In Table 3, we list the real parts and imaginary parts of $ S_0(x,0) $
and $ S_5(x,0) $ for the second order ( $ k = 2 $ ) fermion propagator,
along the diagonal ( $ x_1 = x_2 $ ) of the $ 16 \times 16 $ lattice.

{\footnotesize
\begin{table}
\begin{center}
\begin{tabular}{|c|c|c|c|c|c|}
\hline
$ x_1 $ & $ x_2 $ &  Re($S_0$) &  Im($S_0$)  &  Re($S_5$)  &  Im($S_5$) \\
\hline
\hline
0.0   &   0.0  &   0.8950  &  0.0  & $ 2.94 \times 10^{-3} $ &  0.00 \\
\hline
1.0   &   1.0  &  -0.0609  &  0.0156 & $  1.42 \times 10^{-3} $
                                     & $ -3.81 \times 10^{-4} $ \\
\hline
2.0   &   2.0  &  $-6.31 \times 10^{-3} $ & $ 5.97 \times 10^{-3} $
               &  $-2.25 \times 10^{-4} $ & $ 1.30 \times 10^{-4} $ \\
\hline
3.0   &   3.0  &  $ 7.24 \times 10^{-4} $ & $ -3.08 \times 10^{-3} $
               &  $-1.08 \times 10^{-4} $ & $  3.48 \times 10^{-4} $ \\
\hline
4.0   &   4.0  &  $-8.25 \times 10^{-4} $ & $ -1.43 \times 10^{-3} $
               &  $ 6.01 \times 10^{-5} $ & $  9.28 \times 10^{-5} $ \\
\hline
5.0   &   5.0  &  $-6.54 \times 10^{-4} $ & $  7.07 \times 10^{-6} $
               &  $ 1.04 \times 10^{-5} $ & $ -3.82 \times 10^{-5} $ \\
\hline
6.0   &   6.0  &  $ 3.84 \times 10^{-4} $ & $ -1.05 \times 10^{-4} $
               &  $-1.42 \times 10^{-5} $ & $ -1.30 \times 10^{-5} $ \\
\hline
7.0   &   7.0  &  $ 9.54 \times 10^{-5} $ & $ -1.33 \times 10^{-4} $
               &  $ 8.16 \times 10^{-6} $ &  0.0                     \\
\hline
8.0   &   8.0  &  $ 2.14 \times 10^{-5} $ & $  1.28 \times 10^{-4} $
               & 0.0 & 0.0 \\
\hline
9.0   &   9.0  &  $ 2.27 \times 10^{-4} $ & $  1.36 \times 10^{-4} $
               &  $-6.21 \times 10^{-6} $ & $ -2.31 \times 10^{-6} $  \\
\hline
10.0  &  10.0  &  $ 1.49 \times 10^{-4} $ & $  2.75 \times 10^{-4} $
               &  $-4.21 \times 10^{-6} $ & $ -1.34 \times 10^{-5} $  \\
\hline
11.0  &  11.0  &  $-2.05 \times 10^{-4} $ & $ -7.14 \times 10^{-4} $
               &  $ 2.17 \times 10^{-5} $ & $  5.83 \times 10^{-6} $  \\
\hline
12.0  &  12.0  &  $ 7.16 \times 10^{-4} $ & $ -1.53 \times 10^{-3} $
               &  $-4.10 \times 10^{-5} $ & $  9.00 \times 10^{-5} $  \\
\hline
13.0  &  13.0  &  $ 2.55 \times 10^{-3} $ & $ -1.85 \times 10^{-3} $
               &  $-3.08 \times 10^{-4} $ & $  1.76 \times 10^{-4} $  \\
\hline
14.0  &  14.0  &  $-8.30 \times 10^{-3} $ & $  2.46 \times 10^{-3} $
               &  $-2.70 \times 10^{-4} $ & $  1.27 \times 10^{-5} $  \\
\hline
15.0  &  15.0  &  -0.0629  & $ 8.62 \times 10^{-4} $
               &  $ 1.46 \times 10^{-3} $ & $  -2.99 \times 10^{-5} $  \\
\hline
\end{tabular}
\end{center}
\caption{The scalar part $ S_0(x,0) $ and pseudoscalar part $S_5(x,0)$
in the fermion propagator of the second order ( $ k = 2 $ ) Overlap Dirac
operator, in a background gauge field ( see text for the parameters )
on a $ 16 \times 16 $ lattice with antiperiodic boundary conditions.
One of the end points of the propagator is fixed at the origin, while the
other end point is at one of the sites along the diagonal ( $ x_1 = x_2 $ ). }
\label{table:3}
\end{table}
}

In general, the scalar part $ S_0(x,y) $ and the
pseudoscalar part $ S_5 (x,y) $ in the higher order ( $ k = 1, 2 $ )
fermion propagators seem to be local, especially for near continuum
gauge configurations. However, one cannot exclude the possibility that
they may cause the perturbative instability of the pole of the
fermion propagator \cite{twc98:6b}.
On the other hand, for the Neuberger-Dirac operator,
we are sure that $ S_0(x,y) = a/2 \delta_{x,y} $
and $ S_5(x,y) = 0 $ for any gauge configuration,
as well as the perturbative stability of the pole
of the fermion propagator \cite{twc98:6b}.
So, from this viewpoint, the chiral properties of Neuberger-Dirac operator
are better than those of higher-order Overlap Dirac operators.

\subsection{Axial anomaly}

The axial anomaly of GW Dirac operator $ D $ satisfying
(\ref{eq:gwr}) is \cite{ph98:1,ml98:2}
\bea
\label{eq:ax_gw}
\anxL = a \ \tr [ \gm5 ( R D ) (x,x) ]
\eea
where the trace runs over the Dirac and color space.
Substituting $ R = ( a \gm5 D )^{2k} $ into (\ref{eq:ax_gw}),
we obtain
\bea
\label{eq:ax_Dk}
\anxL = \tr [ ( a \gm5 D )^{2k+1} (x,x) ] \ .
\eea

The sum of the axial anomaly over all sites is equal to
the index of $ D $,
\bea
\label{eq:anomaly_sum}
\sum_x \anxL = n_{-} - n_{+} \ .
\eea
If the index of $ D $ is equal to the topological charge $ Q $
of the gauge background, then the sum of the axial anomaly is equal
to $ Q $. However, it does not necessarily imply that $ \anxL $ would
agree with the topological charge density at each site. This happens
only when $ D $ is local.

Since the higher-order ( $ k > 0 $ ) Overlap Dirac operator
(\ref{eq:Dk}) is also topologically proper ( i.e.,
its index agrees with the background topological charge for any gauge
background satisfying the topological bound ), then it follows that its
axial anomaly would agree with the topological charge density at each
site if $ D $ is local.
( i.e., the gauge configuration satisfies the locality bound
which is more restrictive than the topological bound ).

In the following, we compute the axial anomaly $ \anxL $
in a two-dimensional background $ U(1) $ gauge field, for
the Neuberger-Dirac operator and higher-order ( $ k = 1, 2 $ )
Overlap Dirac operators respectively. We compare them to
the topological charge density $ \bar\rho(x) $ (\ref{eq:rho_bar})
of the gauge background on the torus.

The deviation of the axial anomaly of a lattice Dirac operator
in a gauge background can be measured in terms of
\bea
\label{eq:delta}
\delta = \frac{1}{N_s}
         \sum_x \frac{|\anxL - a^2 \bar\rho(x)|} {a^2|\bar\rho(x)|}
\eea
where $ N_s $ is the total number of sites of the lattice,
and $ \bar\rho(x) $ is the topological charge density inside
the square of area $ a^2 $ centered at $ x $.

In a nontrivial $ U(1) $ gauge background with parameters $ Q = 1 $,
$ h_1 = 0.1 $, $ h_2 = 0.2 $, $ A_1^{(0)} = 0.3 $, $ A_2^{(0)} = 0.4 $
and $ n_1 = n_2 = 1 $ ( as defined in Eqs. (7) and (8) in
Ref. \cite{twc98:4} ) on the $ 12 \times 12 $ lattice with $ a = 1 $,
the axial anomaly $ \anxL $ and its deviation
$ \delta $ are computed for the Neuberger-Dirac operator
and higher-order ( $ k = 1, 2 $ ) Overlap Dirac operators respectively.
The results are :
\beq
\label{eq:delta_2d}
\delta = \left\{  \begin{array}{ll}
 0.110, &   \mbox{Neuberger} \ ,   \\
 0.193, &  k = 1  \ ,   \\
 0.351, &  k = 2  \ .   \\
                  \end{array}
              \right.
\eeq

The relatively large deviations of axial anomaly
in higher-order ( $ k = 1, 2 $ ) Overlap Dirac operators
indicate that they are less localized than the
Neuberger-Dirac operator. And the locality of $ D $ gets
worse as the order $ k $ goes higher
( i.e., $ D_{k=2} $ is less localized than $ D_{k=1} $ ).
We have confirmed this by examining $ | D(x,y) | $
versus $ | x - y | $ explicitly.
This implies that the $ \epsilon $ in the locality bound
\bea
|| \Id - U(p) || < \epsilon, \hspace{4mm} \mbox{for all plaquettes}
\eea
for a higher order ( $ k > 0 $ )
Overlap Dirac operator is more restrictive ( smaller ) than that of
the Neuberger-Dirac operator, and it gets smaller as the order
goes higher.

For all background gauge configurations we have tested,
the Neuberger-Dirac operator always gives
anomaly deviation ( $ \delta $ ) smaller than those of the
higher-order ( $ k = 1, 2 $ ) Overlap Dirac operators.

\subsection{Fermion determinant}

The fermion determinant $ \det(D) $ is proportional to the exponentiation
of the one-loop effective action which is the summation of any number of
external sources interacting with one internal fermion loop.
It is one of the most crucial quantities to be examined in any lattice
fermion formulations. The determinant of $ D $ is the product of all its
eigenvalues
\bea
\det(D) = ( 1 + e^{\text{i} \pi} )^{( n_{+} + n_{-} ) } \det'(D)
\eea
where $ \det'(D) $ is equal to the product of all non-zero eigenvalues.
Since the eigenvalues of $ D $ are either real of come in complex
conjugate pairs, $ \det'(D) $ must be real and positive.
For $ Q = 0 $, then $ n_{+} + n_{-} = 0 $ and $ \det(D) = \det'(D) $.
For $ Q \neq 0 $, then $ n_{+} + n_{-} \neq 0 $ and $ \det(D) = 0 $, but
$ \det'(D) $ still provides important information about the spectrum.
In continuum, exact solutions of fermion determinants in the general
background $ U(1) $ gauge fields on a torus ( $ L_1 \times L_2 $ )
was obtained in Ref. \cite{sachs_wipf}.
In the following we compute $ \det'(D) $ for the Neuberger-Dirac operator
and higher-order ( $ k = 1, 2 $ ) Overlap Dirac operators respectively,
and then compare them with the exact solutions in continuum.
For simplicity, we turn off the harmonic part
( $ h_1 = h_2 = 0 $ ) and the local sinusoidal fluctuations
( $ A_1^{(0)} = A_2^{(0)} = 0 $ ) and examine
the change of $ \det'(D) $ with respect to the topological charge
$ Q $. For such gauge configurations, the exact solution \cite{sachs_wipf}
is
\bea
\det'[ D(Q)] = N \sqrt{ \Bigl( \frac{L_1 L_2 }{2 |Q| } \Bigr )^{|Q|} }
\label{eq:exactD}
\eea
where the normalization constant $ N $ is fixed by
$$
N = \sqrt{ \frac{2}{L_1 L_2 } }
$$
such that $ \det'[D(1)] = 1 $.

In Table 4 and Table 5, the fermion determinants $ \det'(D) $
are listed for $ 8 \times 8 $ and $ 16 \times 16 $ lattice respectively.
It is clear that the Neuberge-Dirac operator always produces results better
than those of the higher-order ( $ k = 1, 2 $ ) Overlap Dirac operators.
The first order ( $ k=1 $ ) Overlap Dirac operator performs 
better than the second order ( $ k = 2 $ ) one.
This is essentially due to the fact that $ D $ becomes less localized
as the order ( $ k $ ) goes higher.

{\footnotesize
\begin{table}
\begin{center}
\begin{tabular}{|c|c|c|c|c|}
\hline
    &  exact &  Neuberger &  $k=1$ &  $k=2$  \\
\hline
 Q  &   $ \det'[D(Q)]_{exact} $  &  $ \det'[D(Q)] $
    &   $ \det'[D(Q)]_{k=1}   $  &  $ \det'[D(Q)]_{k=2} $    \\
\hline
\hline
%
  1  &     1.00000  &    1.00000  &   1.00000  &    1.00000    \\
\hline
  2  &     2.82843  &    2.77348  &   3.01695  &    3.05686    \\
\hline
  3  &     6.15840  &    5.96891  &   7.21380  &    7.51909    \\
\hline
  4  &     11.3137  &    10.7157  &   14.6911  &    15.7450    \\
\hline
  5  &     18.3179  &    16.9340  &   26.1688  &    28.2029    \\
\hline
  6  &     26.8177  &    24.2001  &   41.9768  &    44.4919    \\
\hline
  7  &     36.1083  &    32.0006  &   62.1673  &    63.6961    \\
\hline
  8  &     45.2548  &    40.2920  &   86.8790  &    85.2556    \\
\hline
  9  &     53.2732  &    45.7353  &   112.877  &    106.173    \\
\hline
 10  &     59.3164  &    50.2816  &   141.546  &    128.122    \\
\hline
\end{tabular}
\end{center}
\caption{The fermion determinant versus the topological charge $ Q $.
The normalization constant is chosen such that $ \det'[D(1)] = 1 $.
The results on the $ 8 \times 8 $ lattice are listed in the last
three columns for the Neuberger-Dirac operator and higher-order
( $ k = 1, 2 $ ) Overlap Dirac operators respectively.
The exact solutions on the $ 8 \times 8 $ torus are computed
according to Eq. (\ref{eq:exactD}). }
\label{table:4}
\end{table}
}

{\footnotesize
\begin{table}
\begin{center}
\begin{tabular}{|c|c|c|c|c|}
\hline
    & exact  &  Neuberger   &  $ k=1 $  &  $ k=2 $  \\
\hline
 Q  &   $ \det'[D(Q)]_{exact} $  &  $ \det'[D(Q)] $
    &   $ \det'[D(Q)]_{k=1}   $  &  $ \det'[D(Q)]_{k=2} $    \\
\hline
\hline
  1  &     1.00000  &     1.00000  &   1.00000  &  1.00000   \\
\hline
  2  &     5.65685  &     5.66186  &   5.76472  &  5.78270  \\
\hline
  3  &     24.6336  &     24.0615  &   25.8620  &  26.1218 \\
\hline
  4  &     90.5097  &     90.4894  &   98.9598  &  100.627 \\
\hline
  5  &     293.086  &     286.003  &   337.077  &  346.382 \\
\hline
  6  &     858.166  &     822.664  &   1046.70  &  1092.03 \\
\hline
  7  &     2310.93  &     2170.94  &   3018.17  &  3192.52 \\
\hline
  8  &     5792.62  &     5354.28  &   8198.43  &  8786.52 \\
\hline
  9  &     13637.9  &     12309.5  &   20911.5  &  23111.2 \\
\hline
 10  &     30370.0  &     27336.6  &   51335.5  &  57563.2 \\
\hline
\end{tabular}
\end{center}
\caption{The fermion determinant versus the topological charge
$ Q $ on the $ 16 \times 16 $ lattice.
The normalization constant is chosen such that $ \det'[D(1)] = 1 $.
The results on the lattice are listed in the last
three columns for the Neuberger-Dirac operator and higher-order
($ k = 1, 2 $) Overlap Dirac operators respectively.
The exact solutions on the $ 16 \times 16 $ torus
are computed according to Eq. (\ref{eq:exactD}). }
\label{table:5}
\end{table}
}

\section{ Discussions and Conclusions }

We can understand the emergence of Fujikawa's proposal
by the following considerations.

If one requires that $ D $ is $\gm5$-hermitian,
\BAN
D^{\dagger} = \gm5 D \gm5 \ ,
\EAN
and normal,
\BAN
D^{\dagger} D = D D^{\dagger} \ ,
\EAN
( Note that these two conditions are sufficient to guarantee that
  the real eigenmodes of $ D $ have definite chirality (\ref{eq:real})
  and satisfy the chirality sum rule (\ref{eq:chi_sum_rule}), and
  each complex eigenmode has zero chirality (\ref{eq:complex}). ),
then one immediately obtains
\BAN
\gm5 D \gm5 D = D \gm5 D \gm5 \ .
\EAN
Multiplying above equation by $ \gm5 $, we obtain
\BAN
\gm5 \gm5 D \gm5 D = \gm5 D \gm5 D \gm5 \ ,
\EAN
which can be rewritten as
\BAN
\gm5 ( a \gm5 D )^2 = ( a \gm5 D )^2 \gm5 \ .
\EAN
Since $ ( a \gm5 D )^2 $ is Hermitian and commutes with $ \gm5 $,
one finds an example of $ R = ( a \gm5 D )^2 $ which depends on $ D $.
Then it is straightforward to generalize this $ R $ to any powers,
\BAN
R = ( a \gm5 D )^{2k}, \hspace{4mm} k=0,1,2, \cdots
\EAN
Substituting this $ R $ into the GW relation (\ref{eq:gwr}), we obtain
(\ref{eq:gwr_f}) which is equivalent to Fujikawa's proposal
(\ref{eq:gwr_of}).

It seems to us that Fujikawa's higher-order realization of the
Overlap Dirac operator may not be feasible for practical
computations in lattice QCD, in view of its locality,
chiral properties and computational accessibility in comparison
with those of the Neuberger-Dirac operator.
Nevertheless, from a theoretical viewpoint, it has widened
our scope and deepened our understanding of the Overlap which
does capture one of the fundamental aspects of the nature.

\eject

\bigskip
\bigskip
\flushpar
{\bf Acknowledgement }
\bigskip

\noindent
This work was supported by the National Science Council, R.O.C.
under the grant number NSC89-2112-M002-017. The motivation of this
work emerged after a visit to the Center for Subatomic Structure
of Matter ( CSSM ) at University of Adelaide, where I had
stimulating discussions with Kazuo Fujikawa.
I also thank David Adams, Kazuo Fujikawa, Urs. Heller and
Tony Williams for interesting discussions at CSSM,
and I am grateful to David Adams and Tony Williams for their
kind hospitality.

\bigskip
\bigskip



\begin{thebibliography}{15}


\bibitem{gwr} P. Ginsparg, K. Wilson, Phys. Rev. D25, 2649 (1982).

\bibitem{rn95} R. Narayanan, H. Neuberger, Nucl. Phys. B 443, 305 (1995).

\bibitem{hn97:7} H. Neuberger, Phys. Lett. B 417, 141 (1998);
Phys. Lett. B427, 353 (1998).

\bibitem{fuji00:4} K. Fujikawa, hep-lat/0004012.

\bibitem{no-go}  H.B. Nielsen, N. Ninomiya, Nucl. Phys. B 185, 20 (1981)
[ E: {\it ibid} B195, 541 (1982) ]; {\it ibid} B193, 173 (1981).

\bibitem{twc98:4} T.W. Chiu, Phys. Rev. D58, 074511 (1998).

\bibitem{twc99:11} T.W. Chiu, hep-lat/9911010.

\bibitem{twc98:10a} T.W. Chiu, Phys. Rev. D60, 114510 (1999).

\bibitem{twc98:6b} T.W. Chiu, C.W. Wang, S.V. Zenkin,
Phys. Lett. B 438 (1998) 321.

\bibitem{ph98:1} P. Hasenfratz, V. Laliena, F. Neidermayer,
Phys. Lett. B 427, 125 (1998).

\bibitem{ml98:2} M. L\"uscher, Phys. Lett. B428, 342 (1998).

\bibitem{sachs_wipf} Sachs and Wipf, Helv. Phys. Acta 65 (1992) 652.

\end{thebibliography}
\end{document}